\documentclass[a4paper, authoryear, 12pt]{elsarticle}

\usepackage{soul}
\usepackage{graphics}
\usepackage{graphicx}
\usepackage{psfig}
\usepackage{rotating}
\usepackage[absolute]{textpos}
\usepackage{color}
\usepackage{lineno}
\usepackage[algo2e]{algorithm2e} 
\usepackage[utf8]{inputenc}
\usepackage{amsmath}

\usepackage{amsfonts}
\usepackage{amssymb}
\usepackage{amsthm}
\usepackage{mathrsfs}
\usepackage{latexsym}
\usepackage{multirow}
\usepackage{rotating}
\usepackage{caption}
\usepackage{graphicx}
\usepackage{float}
\usepackage{array}
\usepackage{subfigure}
\usepackage{tabularx}
\usepackage{graphicx,psfig}
\usepackage{footnote}
\usepackage[titletoc,toc,title]{appendix}
\usepackage{colortbl}
\usepackage[dvipsnames]{xcolor}
\usepackage{hyperref}
\usepackage{natbib}
\usepackage{amssymb}
\usepackage{soul}
\usepackage{algorithmic}
\usepackage{algorithm}
\usepackage{epsfig}
\usepackage{relsize}
\usepackage{color}
\usepackage{mathtools}
\usepackage{graphicx,booktabs}
\usepackage{multirow}
\usepackage[utf8]{inputenc}
\usepackage[english]{babel}

\newtheorem*{remark}{Remark}

\begin{document}

\title{Real-time forecasts and risk assessment of novel coronavirus (COVID-19) cases: A data-driven analysis}
\author{Tanujit Chakraborty\footnote[1]{\textit{Mail}: tanujit\_r@isical.ac.in}, Indrajit Ghosh\footnote[2]{\textit{Corresponding author}: Indrajit Ghosh (indra7math@gmail.com)}\\
{\scriptsize \textsuperscript{1} SQC and OR Unit, Indian Statistical
Institute, 203, B. T. Road, Kolkata 700108, India}\\
{\scriptsize \textsuperscript{2} AERU, Indian Statistical Institute,
203, B. T. Road, Kolkata 700108, India}}
\begin{abstract}
The coronavirus disease 2019 (COVID-19) has become a public health
emergency of international concern affecting 201 countries and
territories around the globe. As of April 4, 2020, it has caused a
pandemic outbreak with more than 11,16,643 confirmed infections and
more than 59,170 reported deaths worldwide. The main focus of this
paper is two-fold: (a) generating short term (real-time) forecasts
of the future COVID-19 cases for multiple countries; (b) risk
assessment (in terms of case fatality rate) of the novel COVID-19
for some profoundly affected countries by finding various important
demographic characteristics of the countries along with some disease
characteristics. To solve the first problem, we presented a hybrid
approach based on autoregressive integrated moving average model and
Wavelet-based forecasting model that can generate short-term (ten
days ahead) forecasts of the number of daily confirmed cases for
Canada, France, India, South Korea, and the UK. The predictions of
the future outbreak for different countries will be useful for the
effective allocation of health care resources and will act as an
early-warning system for government policymakers. In the second
problem, we applied an optimal regression tree algorithm to find
essential causal variables that significantly affect the case
fatality rates for different countries. This data-driven analysis
will necessarily provide deep insights into the study of early risk
assessments for 50 immensely affected countries.
\end{abstract}


\maketitle
%\end{frontmatter}

\section{Introduction} \label{Introduction}
In December 2019, Wuhan city of China became the centre of an
outbreak of pneumonia of unknown cause, latter named as coronavirus
disease 2019 (COVID-19), which raised intense attention not only
within China but internationally \citep{guan2020clinical,
wang2020novel}. The COVID-19 pandemic is the most significant global crisis since the World War-II that affected almost all the Countries of our planet \citep{boccaletti2020modeling}. As of April 4, 2020, an outbreak of COVID-19 has resulted in 11,16,643 confirmed cases with reported deaths of 59,170
worldwide \citep{nishiura2020serial}. On March 11, WHO publicly
characterized COVID-19 as a ``global pandemic", and shortly after
that, the United States declared COVID-19 outbreaks a national
emergency. The COVID-19 has caused a great threat to the health and
safety of people all over the world due to its widespread and
potential harm. Thus, the studies of the novel COVID-19 epidemics
and its future development trend has become a cutting-edge research
topic at this moment. We are therefore motivated to ask: (a) Can we
generate real-time forecasts of daily new COVID-19 cases for
countries like Canada, France, India, South Korea, and the UK? (b)
What are the probable causal variables that have significant impacts
on the case fatality rates for the profoundly affected countries?

To answer the first question, we study classical and modern
forecasting techniques for which the prediction accuracy largely
depend on the availability of data
\citep{petropoulos2020forecasting}. In outbreaks of COVID-19
epidemics, there are limited data available, making predictions
widely uncertain. From previous studies, it was evident that the
timing and location of the outbreak facilitated the rapid
transmission of the virus within a highly mobile population
\citep{roosa2020real}. In most of the affected countries, the
governments implemented a strict lockdown in subsequent days of
initial transmission of the virus and within hospitals, patients who
fulfill clinical and epidemiological characteristics of COVID-19 are
immediately isolated. The constant increase in the global number of
COVID-19 cases is putting a substantial burden on the health care
system for Canada, France, India, South Korea, and the UK. To
anticipate additional resources to combat the epidemic, various
mathematical and statistical forecasting tools \citep{li2020early,
wu2020nowcasting} and outside China \citep{kucharski2020early,
zhuang2020preliminary, fanelli2020analysis} were applied to generate
short-term and long-term forecasts of reported cases. These model
predictions have shown a wide range of variations. Since the time
series datasets of COVID-19 contain both nonlinear and nonstationary
patterns, therefore, making decisions based on an individual model
would be critical. In this study, we propose a hybrid modeling
approach to generate short-term forecasts for multiple countries. In
traditional time series forecasting, the autoregressive integrated
moving average (ARIMA) model is used predominantly for forecasting
linear time series \citep{box2015time}. But in recent literature,
the wavelet transformation based forecasting model has shown
excellent performance in nonstationary time series data modeling
\citep{percival2000wavelet}. Thus, combining both models may
accurately model such complex autocorrelation structures in the
COVID-19 time-series datasets and reduce the bias and variances of
the prediction error of the component models. In the absence of
vaccines or antiviral drugs for COVID-19, these estimates will
provide an insight into the resource allocations for the exceedingly
affected countries to keep this epidemic under control. Besides
shedding light on the dynamics of COVID-19 spreading, the practical
intent of this data-driven analysis is to provide government
officials with realistic estimates for the magnitude of the epidemic
for policy-making.

The second problem is connected with the global concern of health
and mortality due to the significant COVID-19 outbreaks. Mortality is
crudely estimated using a statistic, the case fatality rate (CFR),
which divides the number of known deaths by the total number of
identified cases \citep{jung2020real, boldog2020risk,
russell2020estimating}. During the current phase of this global
pandemic, it is critically important to obtain reliable estimates of
the overall CFR. The estimates of CFR are highly dependent on
several country-specific demographic parameters and various disease characteristics. A key differentiation among the CFR of different countries can be found by determining an exhaustive list of causal variables that significantly affect CFR. In this work, we put an
effort to identify critical parameters that may help to assess the
risk (in terms of CFR) using an optimal regression tree model
\citep{breiman1984classification}. The regression tree has a
built-in variable selection mechanism from high dimensional variable
space and can model arbitrary decision boundaries. The regression tree combines case estimates, epidemiological characteristics of the
disease, and heath-care facilities to assess the risks of major
outbreaks for profoundly affected countries. Such assessments will
help to anticipate the expected morbidity and mortality due to
COVID-19 and provide some critical information for the planning of
health care systems in various countries facing this epidemic.

The rest of the paper is organized as follows. In Section
\ref{section_2}, we discuss the data, development of the hybrid
model, and experimental results for short-term forecasts of COVID-19
for Canada, France, India, South Korea, and the UK. In Section
\ref{section_3}, country-wise CFR datasets, method, and results for
finding critical parameters are presented. We discuss the
assumptions and limitations of our findings in Section
\ref{section_4}. Finally, the discussions about the results and
policy recommendations are given in Section \ref{section_5}.

\section{Real-time forecasting of COVID-19 cases}\label{section_2}

We focus on the daily figures of confirmed cases for five different
countries, namely Canada, France, India, South Korea, and the UK.
The datasets are retrieved by the Global Change Data
Lab\footnote{\url{https://ourworldindata.org/coronavirus}}). All
these datasets are collected from the starting date of the disease
for the respective countries to April 4, 2020. In this section, we
first briefly discuss these datasets, followed by the development of
the proposed hybrid model, and finally, the application of the proposed model to generate short-term forecasts of the future COVID-19 cases for five different countries. All these datasets and codes to be used in this section are made publicly available at \url{https://github.com/indrajitg-r/COVID} for the reproducibility of this work.

\subsection{Datasets}

Five univariate time-series datasets are collected for the real-time
prediction purpose of COVID-19 cases for India, Canada, France,
South Korea, and the UK. Several previous studies have forecasted
future COVID cases for China and a few other countries using
mathematical and traditional time series forecasting models, for
details see \citep{roosa2020real, li2020early, wu2020nowcasting,
kucharski2020early, zhuang2020preliminary}. We try to nowcast the
COVID-19 cases of five different countries based on their past
cases. For India and UK, we consider the daily laboratory-confirmed
cases from January 30, 2020, through April 4, 2020 and from January 31, 2020 through April 4, 2020, respectively, for model building. Daily
COVID-19 cases data for Canada, France, and South Korea are taken
for the time period January 20, 2020 through April 4, 2020, January 25, 2020 through April 4 2020, and January 26 through April 4 2020,
respectively. 

The dataset for India contains a total of 64 observations, 65 observations for the UK, 70 observations for Canada, 71 observations for France, and 76 for South Korea. For these five countries the outbreaks of COVID-19 started almost from the same timeline and the epidemic curves still not showing the sharp diminishing nature, just like China. We limit our attention to trended and non-seasonal models, given the patterns, observed in Table 1.%\ref{table11}. 
Note that we follow a pragmatic approach in that we assume that the trend will continue indefinitely in the future in contradiction with other S-curve or deterministic SIR modeling approaches which assume convergence.

\begin{figure}[H]
\centering
\includegraphics[width=0.9\textwidth]{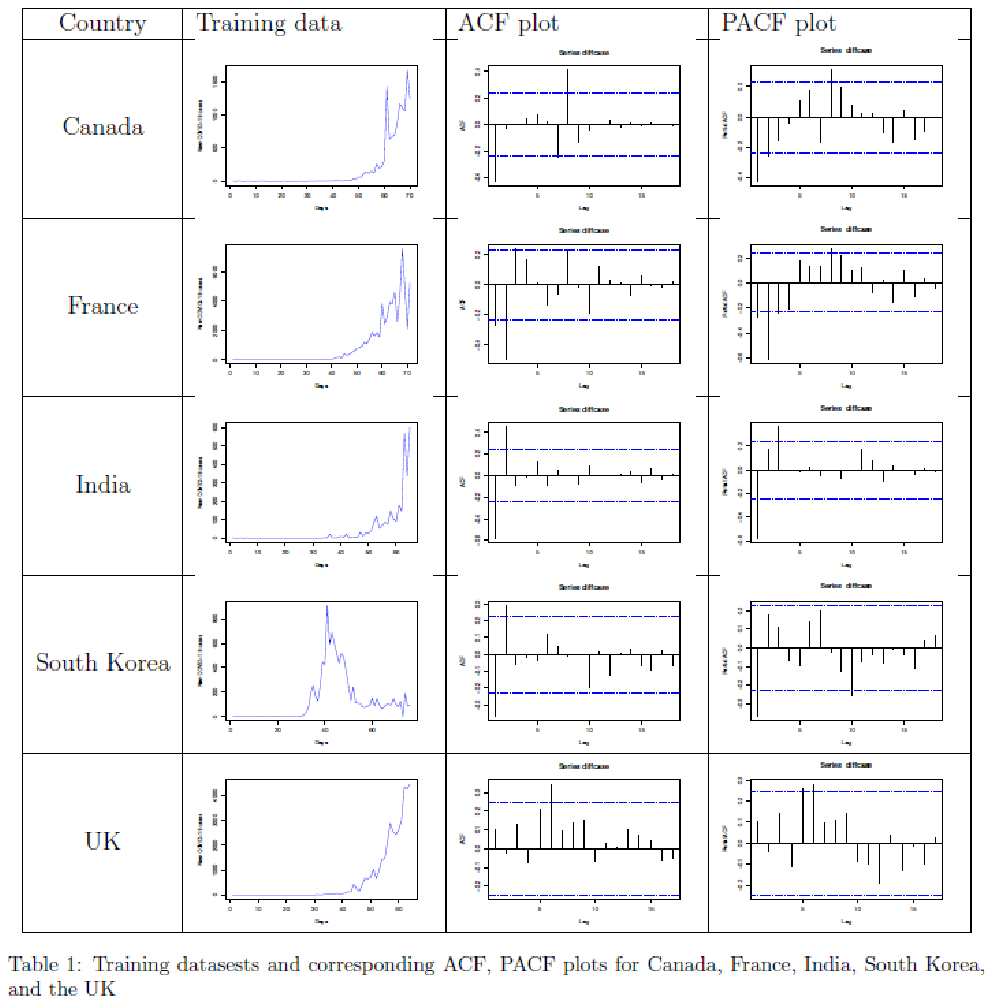}
\end{figure}

\subsection{Proposed Model}

To forecast confirmed cases of COVID-19, we adopt hybrid time series
forecasting approaches combining ARIMA and wavelet-based forecasting techniques. The proposed hybrid model overcome the deficiencies of the single time series models. Before describing the proposed methodology, we give a brief description of the individual models to be used in the hybridization.

\subsubsection{ARIMA Model}\label{arima}

ARIMA is a classical time series model, used for tracking linear
tendencies in stationary time series data. ARIMA model is denoted by ARIMA($p,d,q$). The parameters $p$ and $q$ are the order of the AR model and the MA model respectively, and $d$ is the level of
differencing \citep{chatfield2016analysis}. ARIMA model can be
mathematically expressed as follows:
\begin{eqnarray*}
{y}_t & = & \theta_0  + \phi_1y_{t-1} + \phi_2y_{t-2} + \cdots +
\phi_py_{t-p} + \varepsilon_{t}   - \theta_1\varepsilon_{t-1} -
\theta_2\varepsilon_{t-2} - \cdots -
          \theta_q\varepsilon_{t-q},
\end{eqnarray*}
where $y_t$ denotes the actual value of the variable under
consideration at time $t$, $\varepsilon_{t}$ is the random error at
time $t$. The $\phi_i$ and $\theta_j$ are the coefficients of the
ARIMA model. The basic assumption made by the ARIMA model is that the error series follows zero mean with constant variance, and satisfies the i.i.d condition. Building an ARIMA model for any given time series dataset can be described in three iterative steps: model
identification (achieving stationarity), parameter estimation (the
autocorrelation function (ACF) and the partial autocorrelation
function (PACF) plots are used to select the values of parameters
$p$ and $q$), and model diagnostics checking (finding the `best'
fitted forecasting model using Akaike Information Criterion (AIC)
and the Bayesian Information Criterion (BIC))
\citep{hyndman2018forecasting}.

\subsubsection{Wavelet-based Forecasting (WBF) Model}\label{wavelet}

Wavelet analysis is a mathematical tool that can
reveal information within the signals in both the time and scale
(frequency) domains \citep{percival2000wavelet}. This property
overcomes the basic drawback of Fourier analysis and wavelet
transforms the original signal data (especially in the time domain)
into a different domain for data analysis and processing.
Wavelet-based models are most suitable for nonstationary data,
unlike ARIMA \citep{nury2017comparative}. Most epidemic and climatic time-series datasets are nonstationary; therefore, wavelet
transforms are used as a forecasting model for these datasets
\citep{fay2007wavelet, aminghafari2007forecasting}. When conducting
wavelet analysis in the context of time series analysis, the
selection of the optimal number of decomposition levels is vital
to determine the performance of the model in the wavelet domain. The following formula for the number of decomposition levels, $W_L = int[log(n)]$ is used to select the number of decomposition levels,
where $n$ is the time-series length. The wavelet-based forecasting (WBF) model transforms the time series data by using a hybrid maximal overlap discrete wavelet transform (MODWT) algorithm with a `haar' filter. Daubechies wavelets can produce identical events across the observed time series in so many fashions that most other time series prediction models cannot recognize \citep{benaouda2006wavelet}. The necessary steps of a wavelet-based forecasting model, defined by \citep{aminghafari2007forecasting}, are as follows. Firstly, the Daubechies wavelet transformation and a decomposition level are applied to the nonstationary time series data. Secondly, the
series is reconstructed by removing the high-frequency component,
using the wavelet denoising method. And, lastly, an appropriate ARIMA model is applied to the reconstructed series to generate
out-of-sample forecasts of the given time series data.

\subsubsection{Hybrid ARIMA-WBF Model}

For the COVID-19 datasets, we propose a hybridization of stationary
ARIMA and nonstationary WBF model to reduce the individual biases of the component models \citep{oliveira2014ensembles}. The COVID-19 cases datasets for five different countries are complex in nature. Thus, the ARIMA model fails to produce random errors or even nonstationary residual series, evident from Figure \ref{residuals}. The behavior of the residual series generated by ARIMA is mostly oscillatory and periodic; thus, we choose the wavelet function to model
the remaining series. Several hybrid models based on ARIMA and neural
networks are available in the field of time series forecasting; see for
example \citep{zhang2003time, aladag2009forecasting, firmino2014correcting, khashei2018comparative, chakraborty2019forecasting, ordonez2019hybrid}.

\begin{algorithm}[H]
\nl Given a time series of length $n$, input the in-sample (training) COVID-19 daily cases data.
\nl Determine the best ARIMA$(p,d,q)$ model using the in-sample (training) data.
\begin{itemize}
 \item ARIMA parameters $p$, $d$, and $q$ values are selected using the procedures described in Section \ref{arima}.
 \item Obtain the predictions using the selected ARIMA$(p,d,q)$ model for the in-sample data and generate required number of out-of-sample forecasts.
 \item Obtain the residual series $(\varepsilon_{t})$ by subtracting ARIMA predicted values from the original training series.
\end{itemize}
\nl Train the residual series $(\varepsilon_{t})$ generated by ARIMA by the WBF model, as described in Section \ref{wavelet}.
\begin{itemize}
\item Select the number of decomposition level using the formulae $W_L = int[log(n)]$ and boundary is chosen to be `periodic'.
\item Obtain in-sample predictions $(\hat{\varepsilon_{t}})$ using the WBF model and generate required number of out-of-sample forecasts..
\end{itemize}
\nl Final predictions $(\hat Y_t)$ are obtained by combining then
ARIMA predictions with WBF predictions $(\hat{\varepsilon_{t}})$ for both the training series as well as the out-of-sample forecasts.
\caption{{\bf Proposed Hybrid ARIMA-WBF Model}} \label{algo1}
\end{algorithm}

Motivated by the above discussion, we propose a novel hybrid
ARIMA-WBF model which is a two-step pipeline approach. In the first step of the proposed hybrid approach, an ARIMA model is built to model the linear components of the epidemic time series, and a set of out-of-sample forecasts are generated. In the second phase, the ARIMA residuals
(oscillatory residual series) are remodeled using a
mathematically-grounded WBF model. Here, WBF models the left-over autocorrelations (in this case, the oscillatory series in Figure \ref{residuals}) in the residuals which ARIMA could not model. The algorithmic presentation of the proposed hybrid
model is given in Algorithm \ref{algo1}.

The proposed model can be looked upon as an error remodeling approach in which we use ARIMA as the base model and remodel its
error series by wavelet-based time series forecasting technique to
generate more accurate forecasts. This is in relevance to model
misspecification in which disturbances in the nonlinear time series of
COVID-19 cases cannot be correctly modeled with the ARIMA model. Therefore, if the error series generated by ARIMA is adequately modeled and incorporated with the forecasts, the performance of the out-of-sample estimates can be improved, even though marginally at
times.

\begin{remark}
The proposed hybrid approach contradicts other mathematical and traditional forecasting modeling approaches applied to COVID-19 data. We choose two completely diverse models for hybridization, one from classical forecasting literature and another from modern forecasting approaches. 
\end{remark}

\subsection{Results}

Five time series COVID-19 datasets for Canada, France, India, South Korea, and the UK are considered for training the proposed model and the component models. The datasets are nonlinear, nonstationary, and non-gaussian in nature. We have used root mean square error (RMSE), mean absolute error (MAE), to evaluate the predictive performance of the models used in this study \citep{james2013introduction}. Since the number of data points in both the datasets is limited thus going for advanced deep learning techniques will simply
over-fit the datasets \citep{hastie2009elements}.

\begin{figure}[H]
\centering
\includegraphics[width=0.28\textwidth]{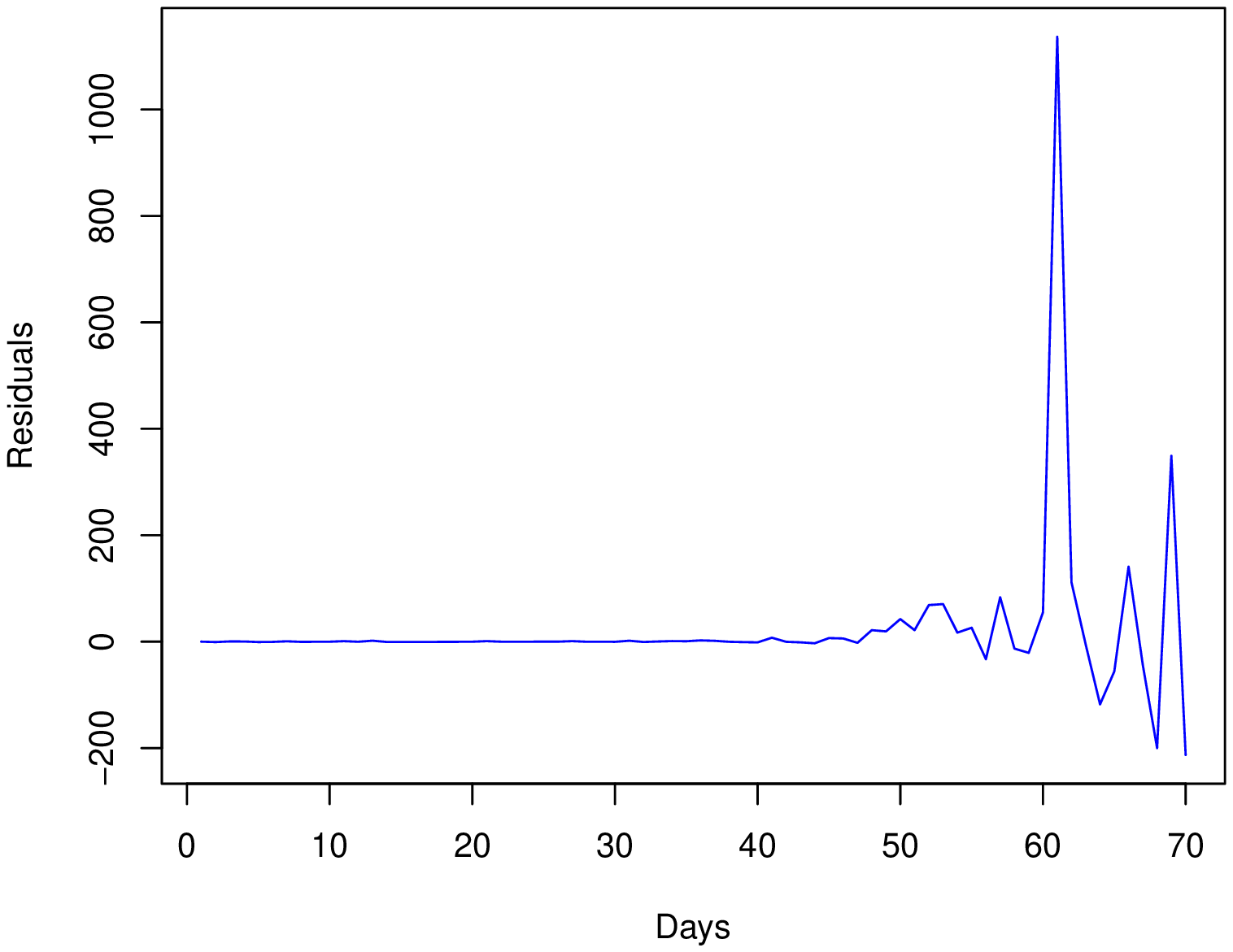}(a)
\includegraphics[width=0.28\textwidth]{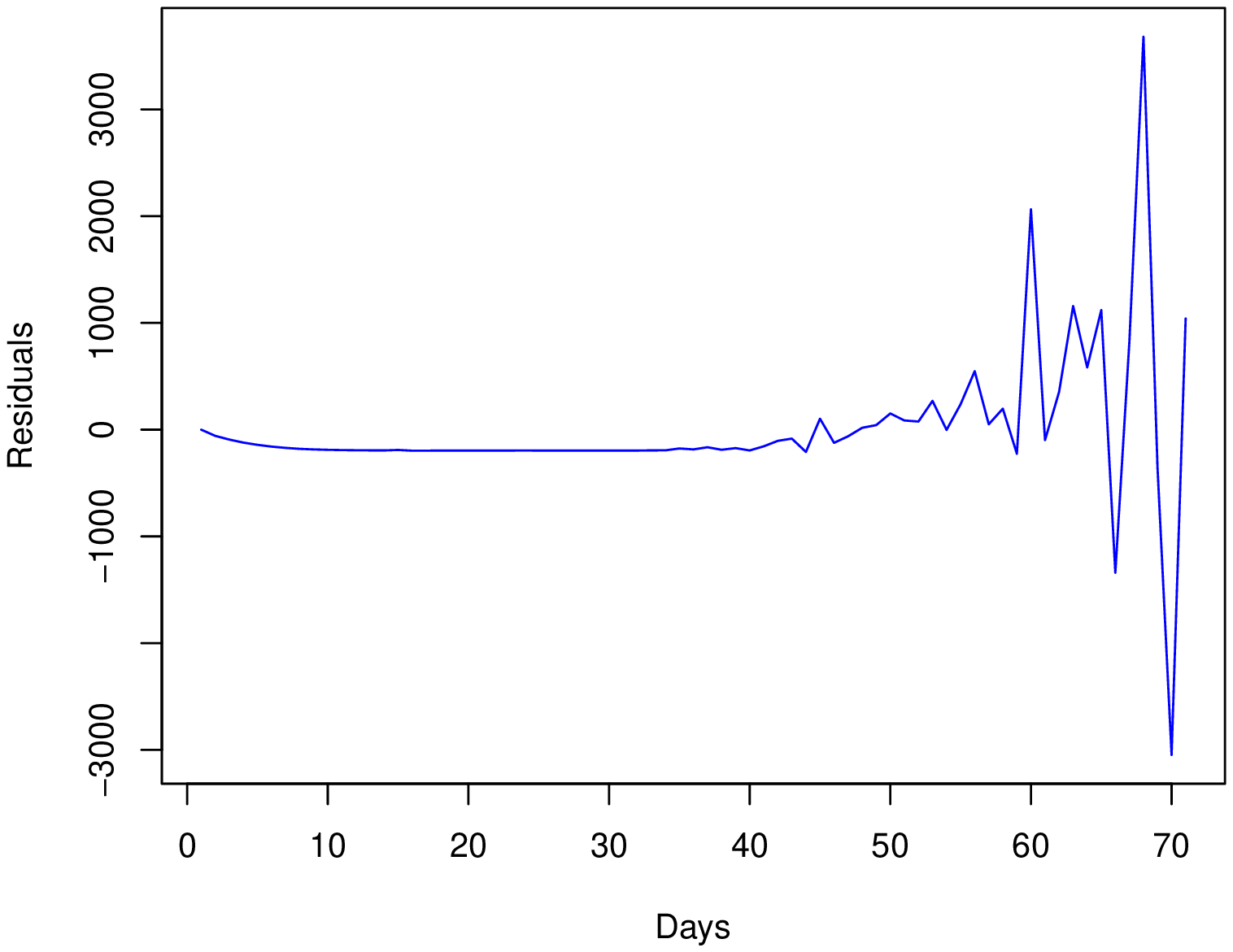}(b)
\includegraphics[width=0.28\textwidth]{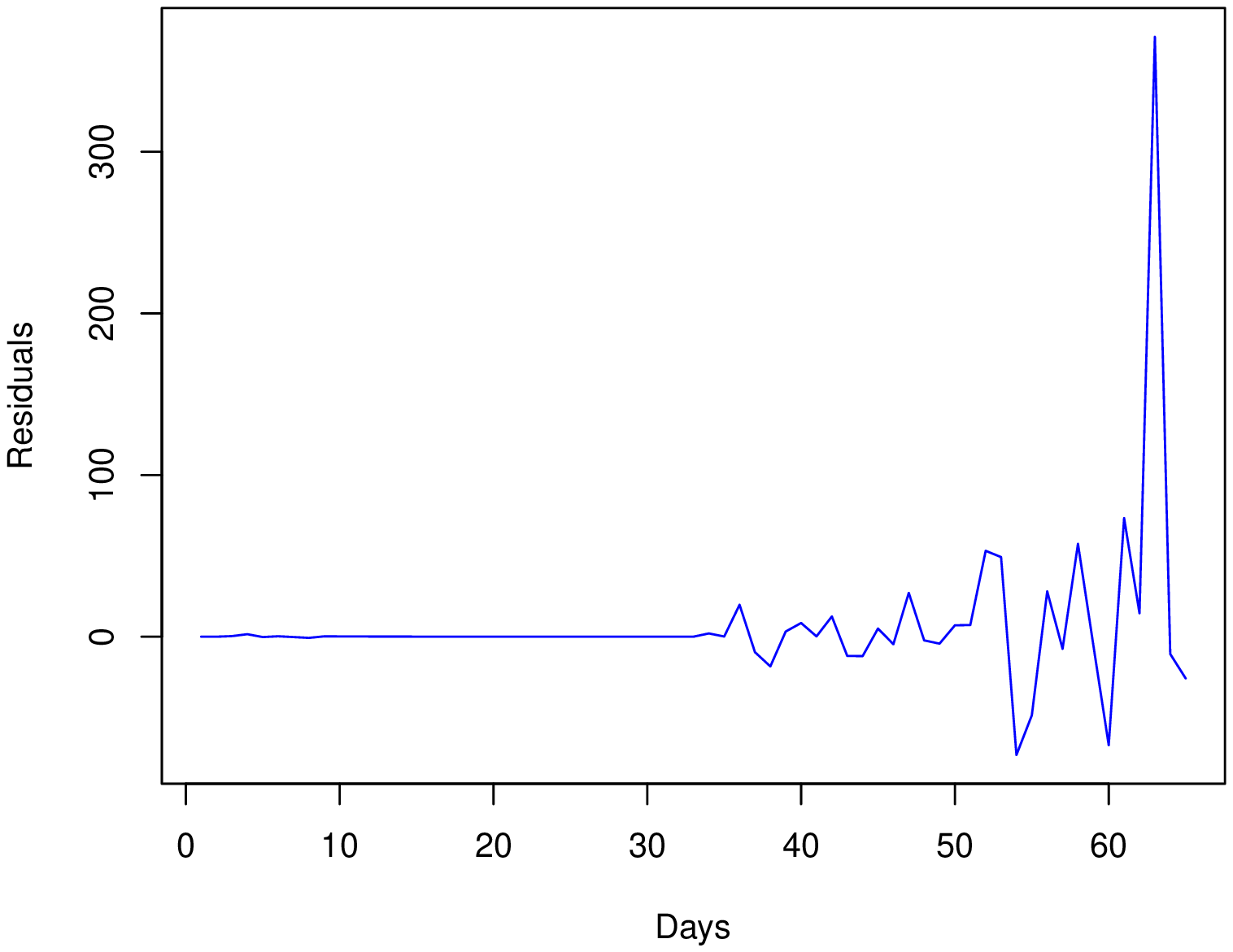}(c)\\
\includegraphics[width=0.28\textwidth]{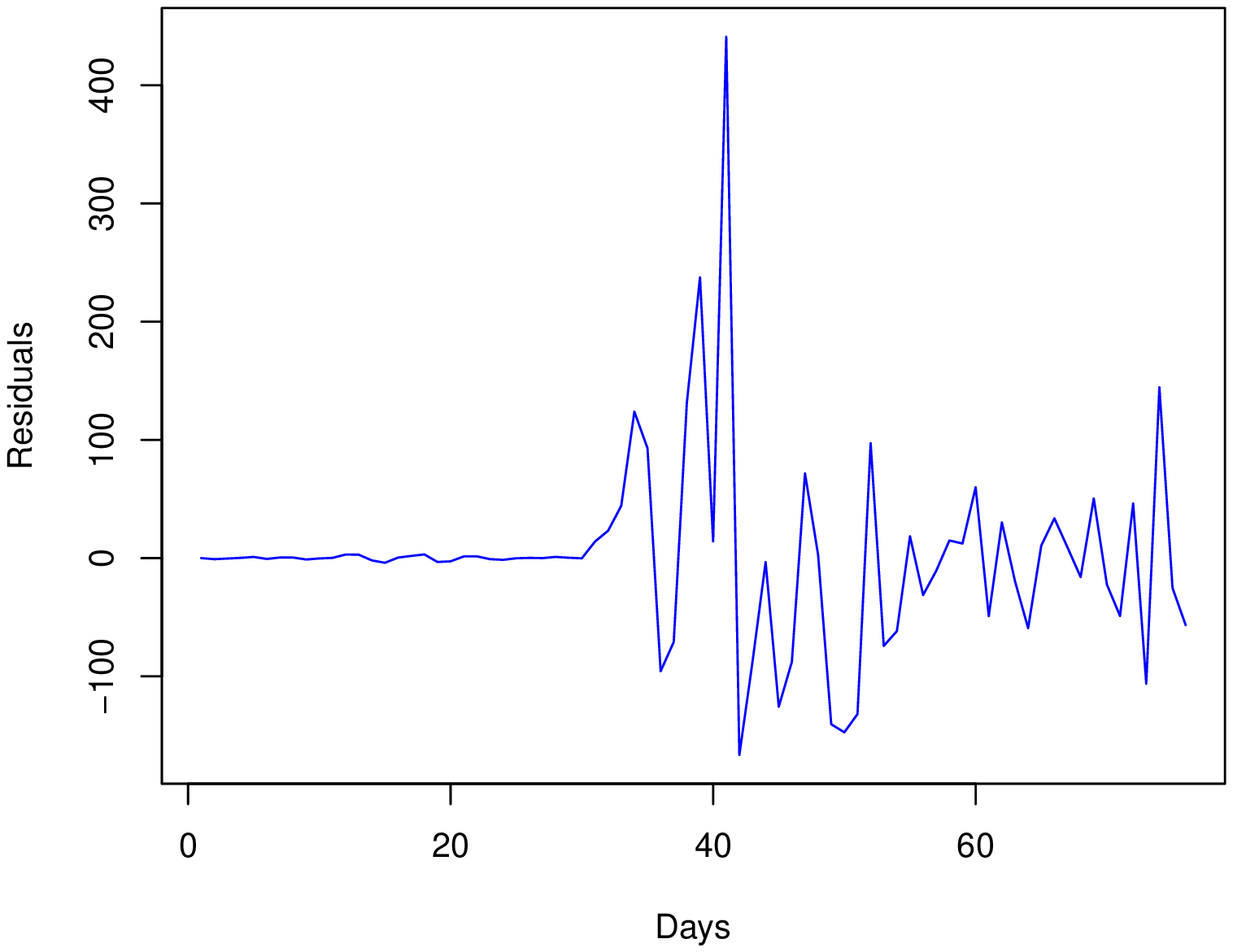}(d)
\includegraphics[width=0.28\textwidth]{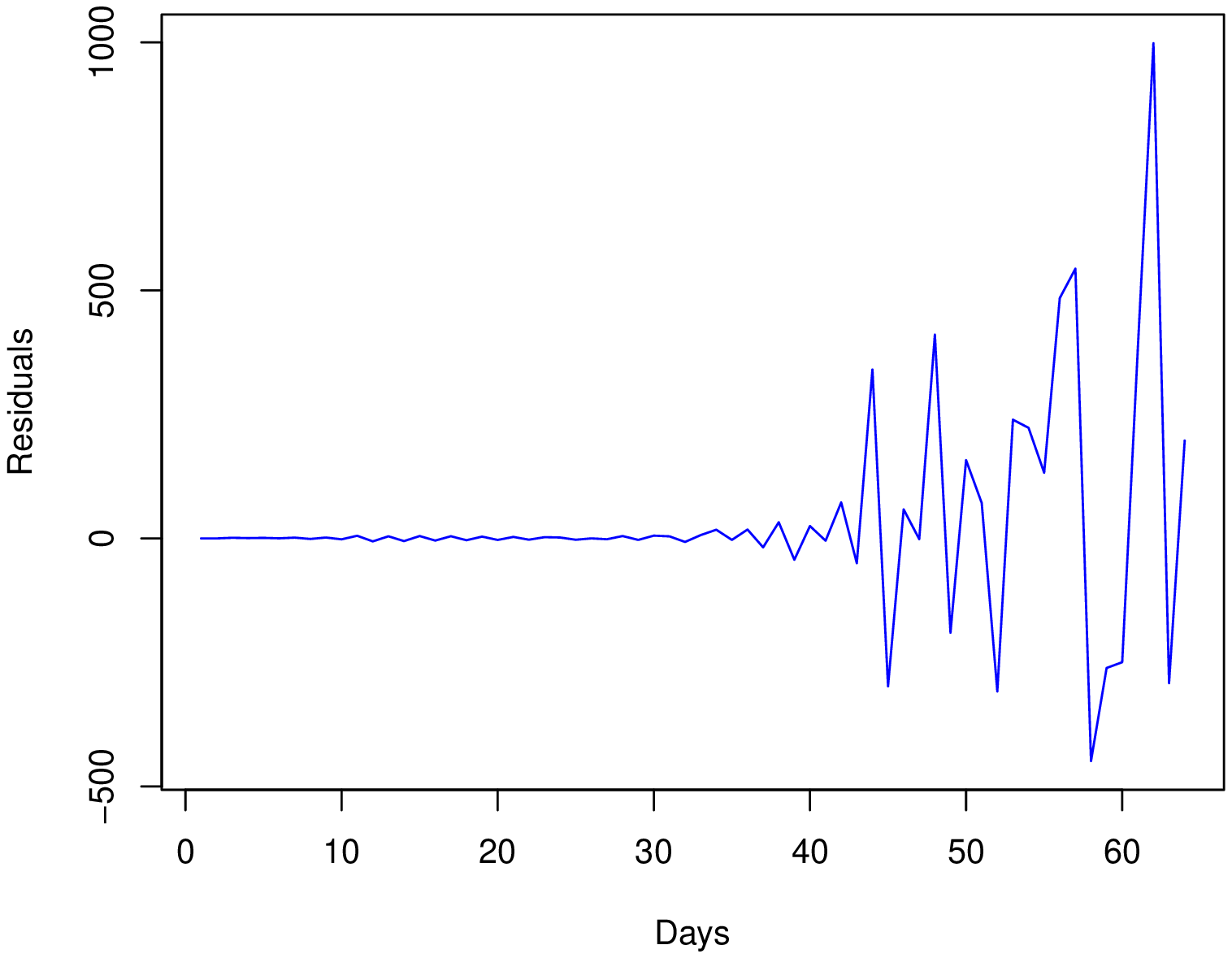}(e)
\caption{Plots of ARIMA residuals for different countries: (a)
Canada; (b) France; (c) India; (d) South Korea; and (e) the
UK.}\label{residuals}
\end{figure}

We start the experimental evaluation for all the five datasets with the classical ARIMA(p,d,q) using `\textit{forecast}' \citep{hyndman2020package} statistical package in R software. To fit an ARIMA model, we first specify the parameters of the model. Using ACF plot and PACF plot (See
Table 1 %\ref{table11}), 
we can decide the value of the parameters of the model. We have also performed unit root tests for stationarity check and all the datasets were found nonstationary. The `best' fitted ARIMA model is chosen using AIC and BIC values for each training dataset. The fitted ARIMA models for five datasets are as follows: ARIMA(1,2,1) for India, ARIMA(1,1,2) for Canada, ARIMA(0,1,1) for France,
ARIMA(2,1,0) for South Korea, and ARIMA(2,2,2) for the UK. We employ a pre-defined Box-Cox transformation set to $\lambda = 0$ to ensure the forecast values stay positive. As the ARIMA model is fitted, forecasts are generated for 10-time steps (5 April 2020 to 14 April 2020) for all the five datasets. We also compute training data predicted values and calculate the residual errors. Plots for the residual series are given in Figure \ref{residuals}. 

\begin{figure}[H]
\centering
\includegraphics[width=0.45\textwidth]{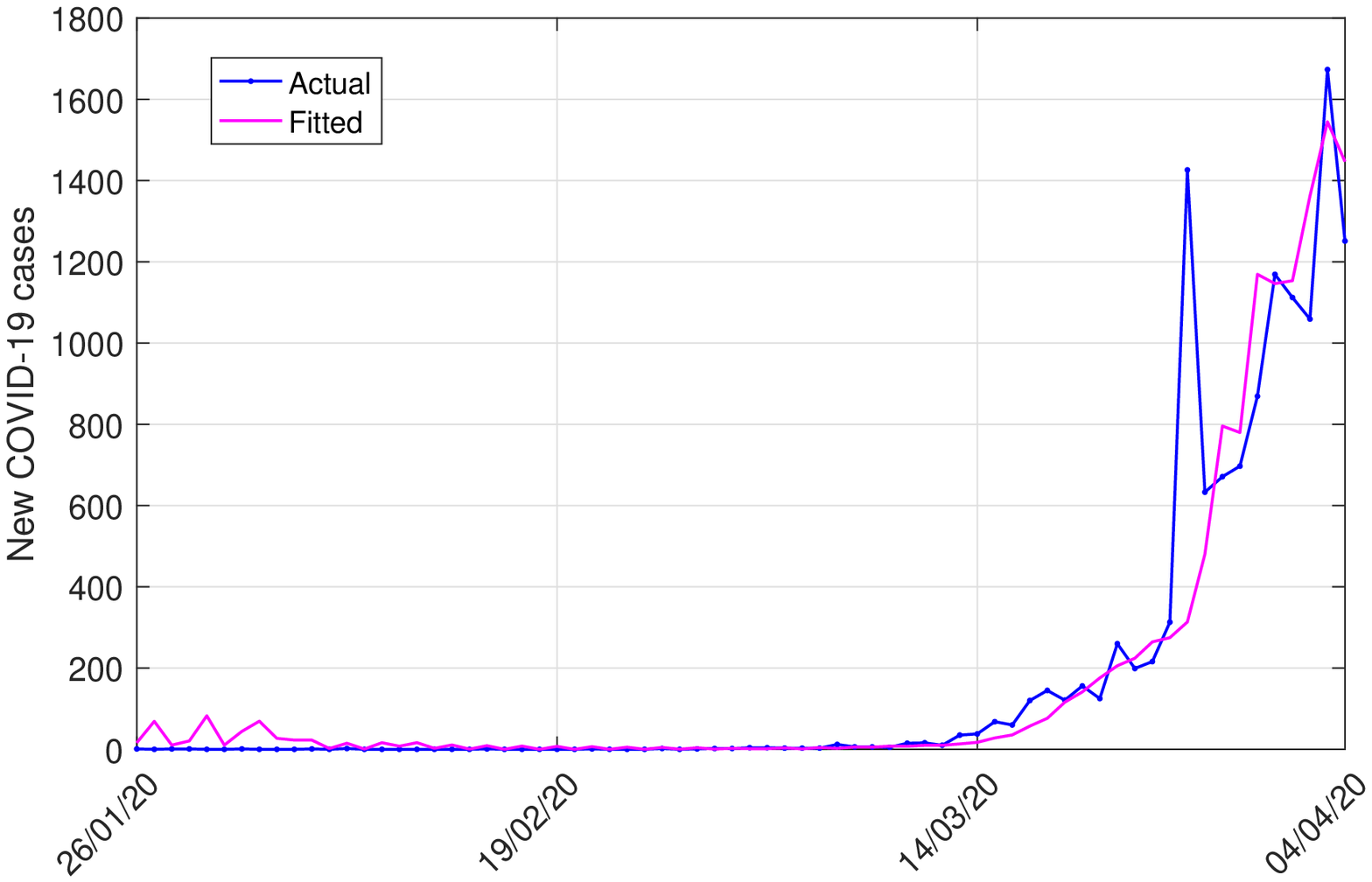}(a)
\includegraphics[width=0.45\textwidth]{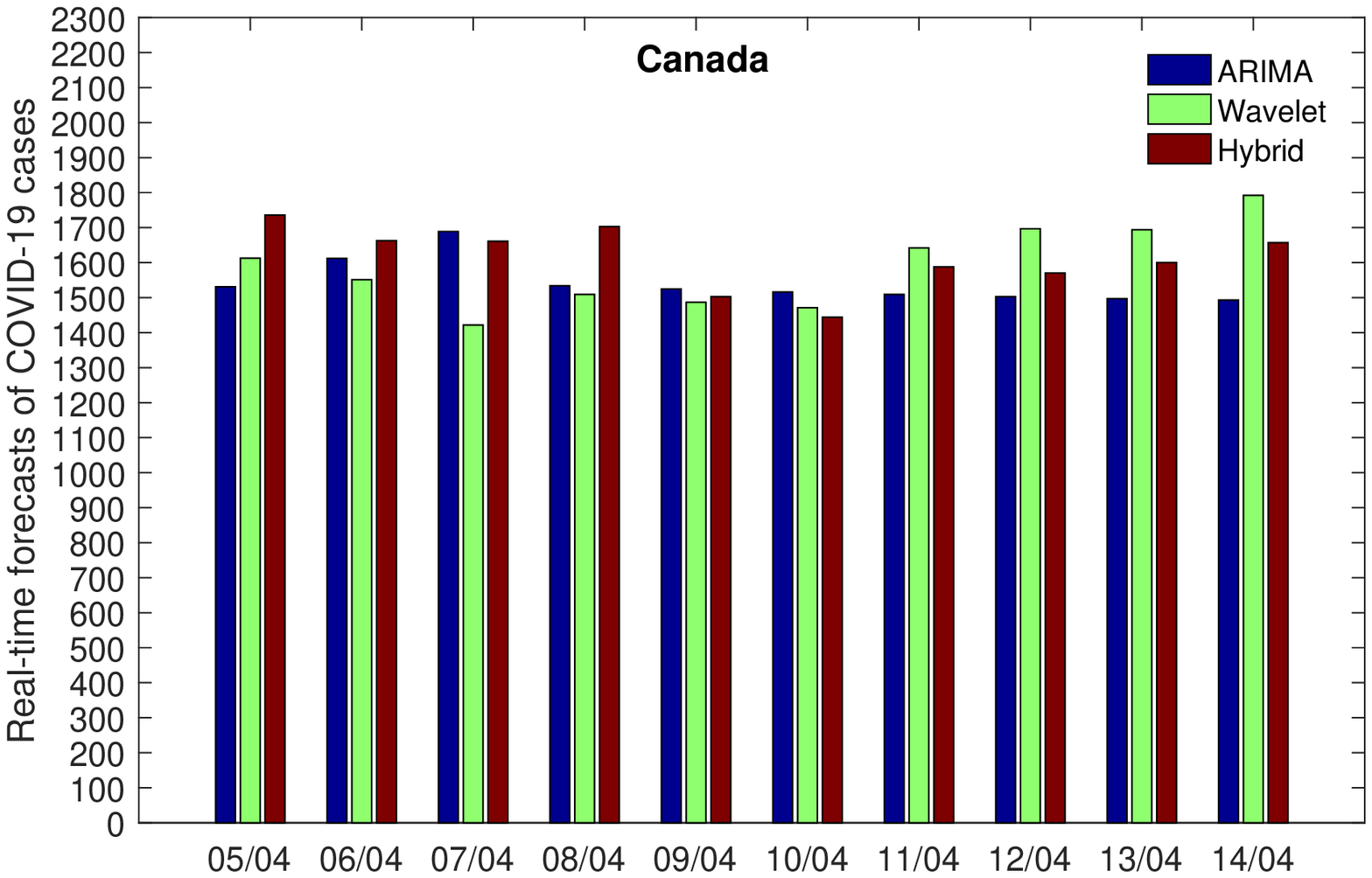}(b)
\caption{Figures of (a) Actual Vs. predicted (Hybrid ARIMA-WBF
Model) values for Canada COVID-19 data; (b) Real-time forecasts (10
days) of the number of cases for Canada}\label{fit_forecast_canada}
\end{figure}

\begin{figure}[H]
\centering
\includegraphics[width=0.45\textwidth]{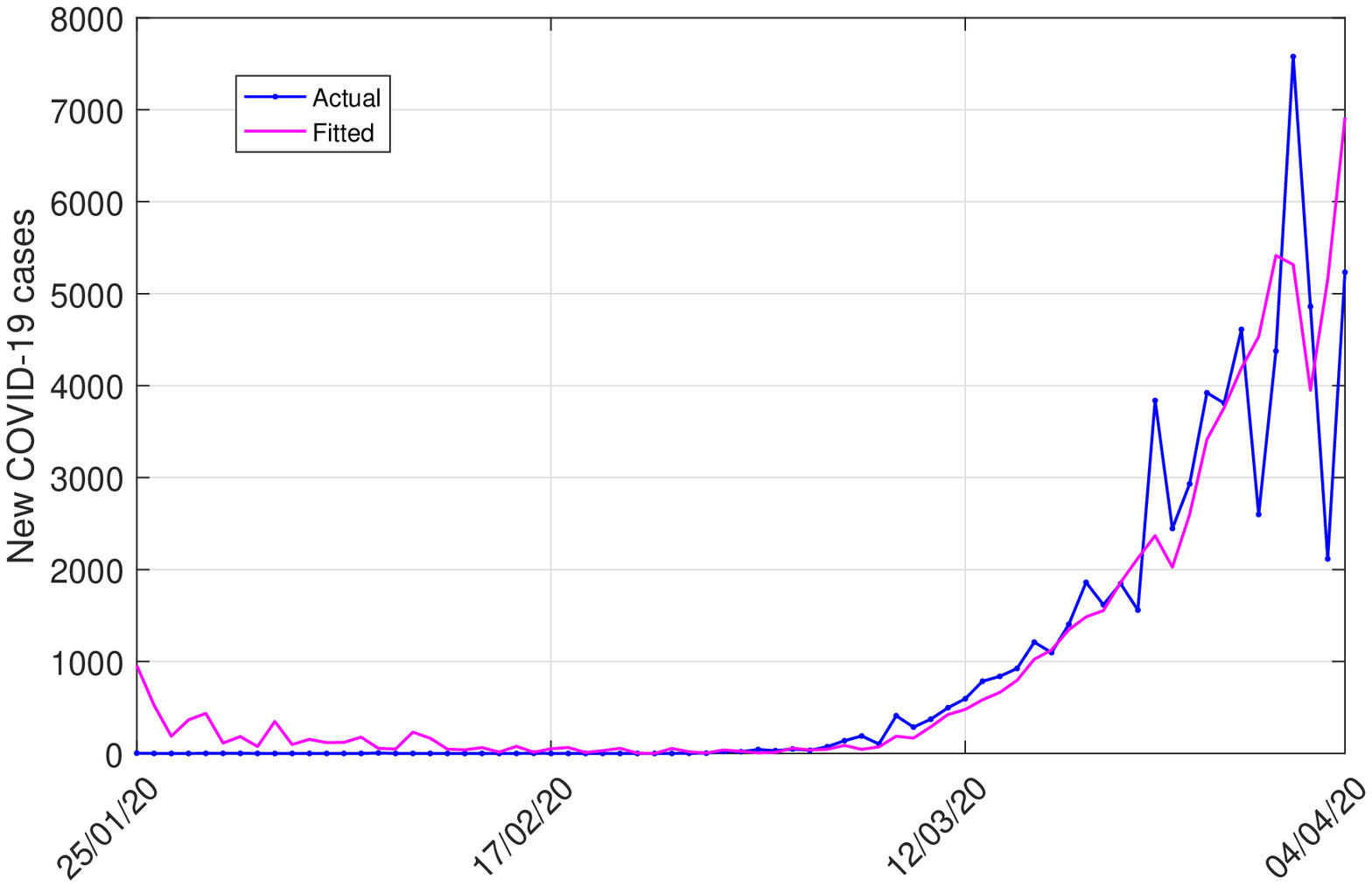}(a)
\includegraphics[width=0.45\textwidth]{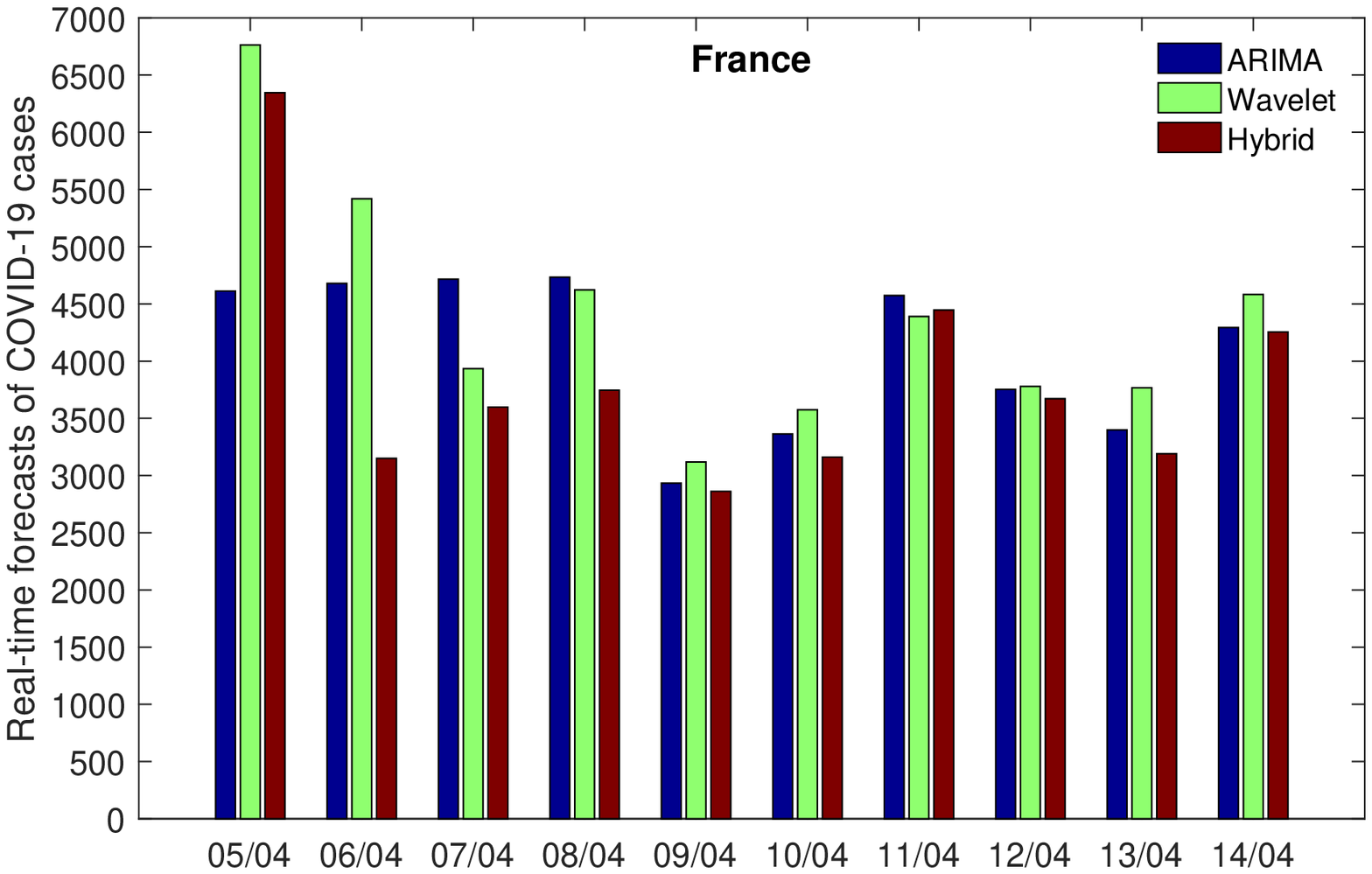}(b)
\caption{Figures of (a) Actual Vs. predicted (Hybrid ARIMA-WBF
Model) values for France COVID-19 data; (b) Real-time forecasts (10
days) of the number of cases for France}\label{fit_forecast_france}
\end{figure}

\begin{figure}[H]
\centering
\includegraphics[width=0.45\textwidth]{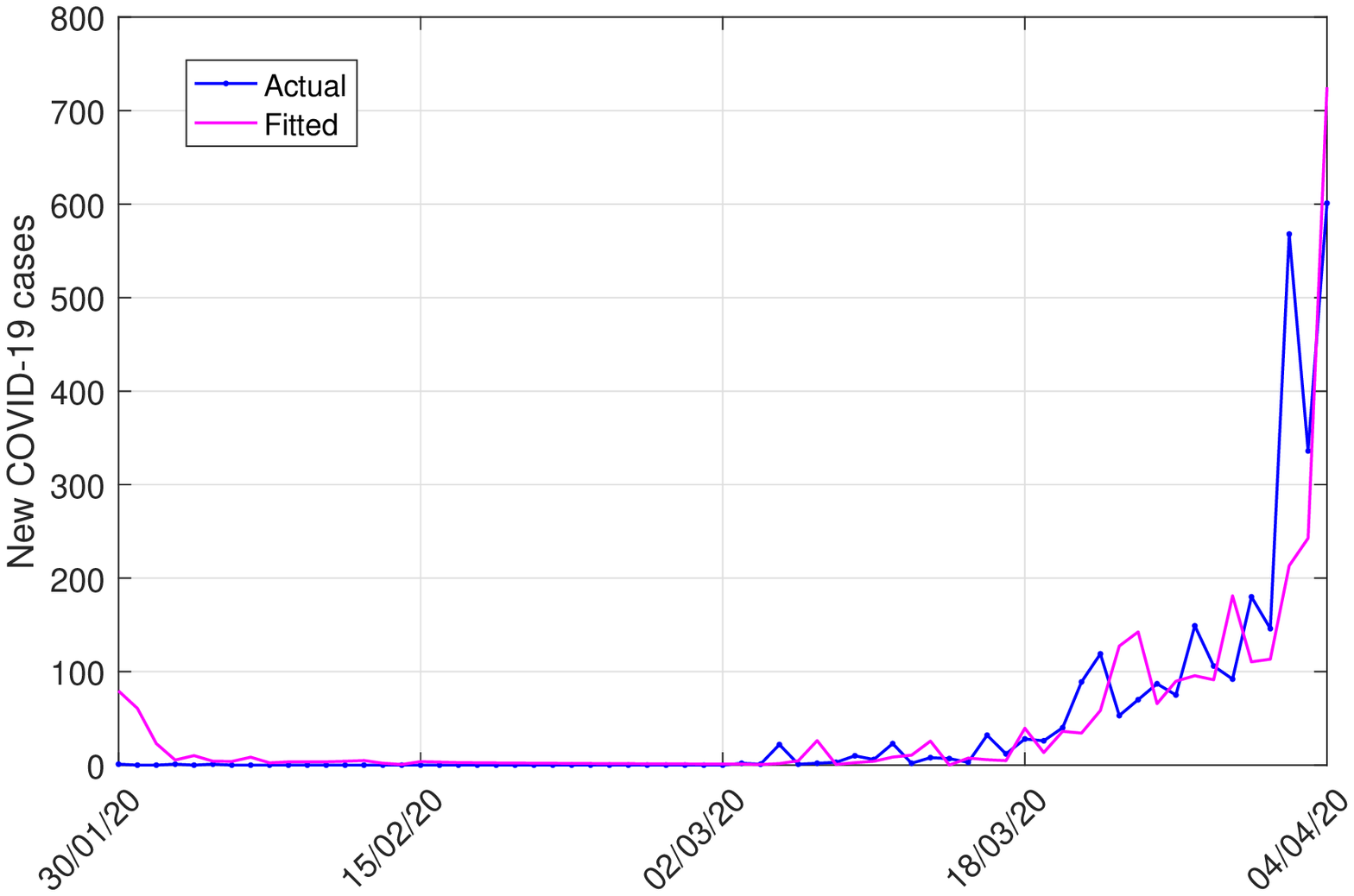}(a)
\includegraphics[width=0.45\textwidth]{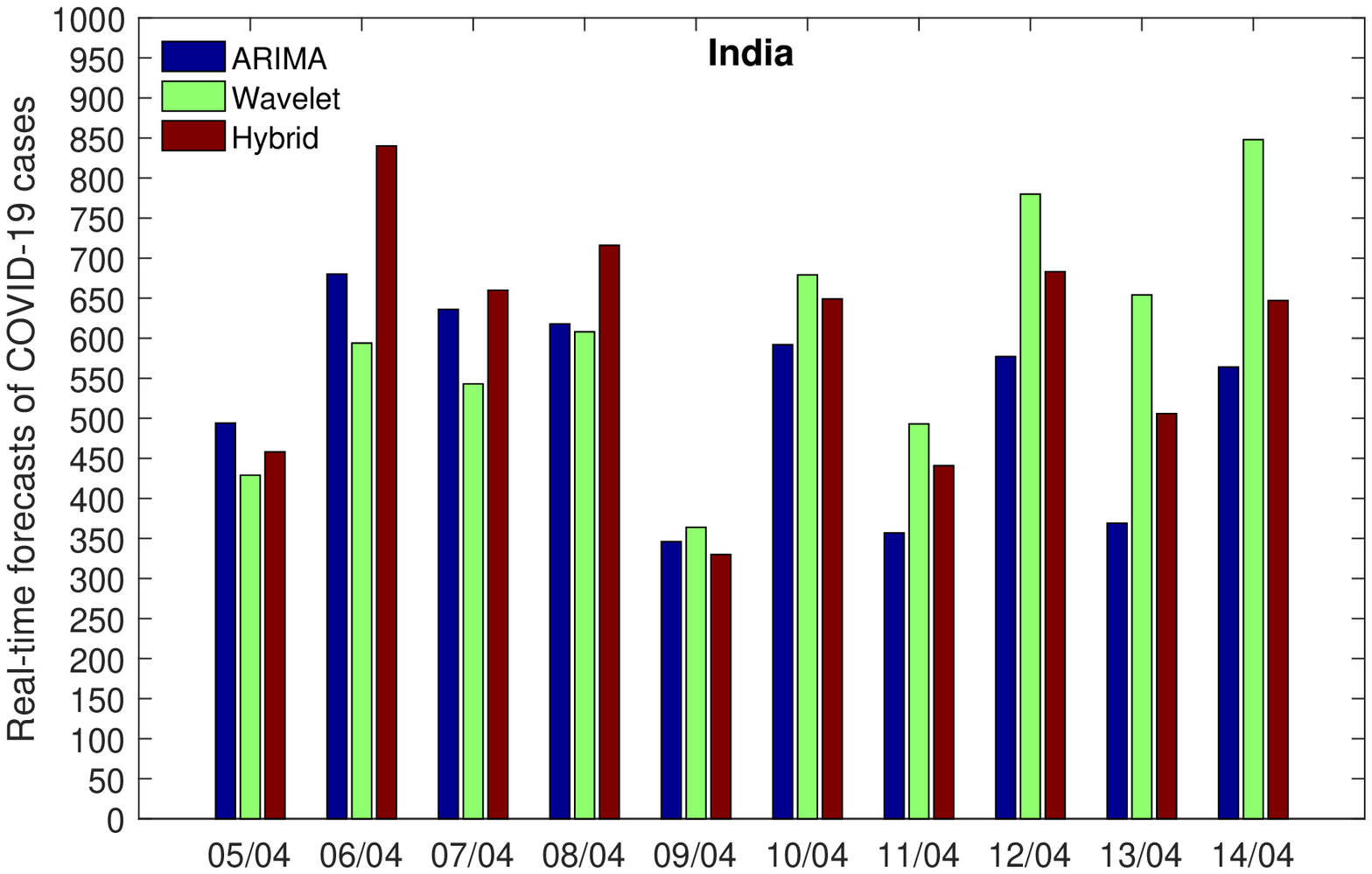}(b)
\caption{Figures of (a) Actual Vs. predicted (Hybrid ARIMA-WBF
Model) values for India COVID-19 data; (b) Real-time forecasts (10
days) of the number of cases for India}\label{fit_forecast_india}
\end{figure}

\begin{figure}[H]
\centering
\includegraphics[width=0.45\textwidth]{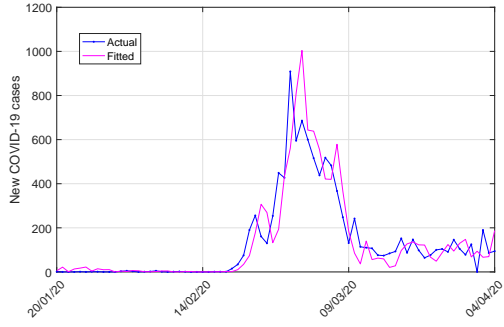}(a)
\includegraphics[width=0.45\textwidth]{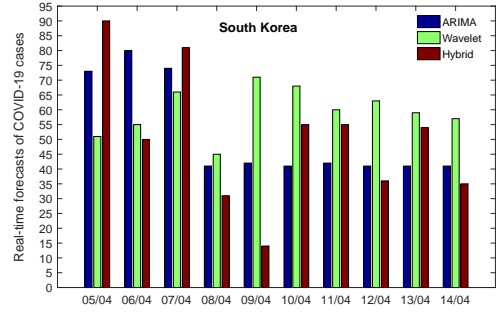}(b)
\caption{Figures of (a) Actual Vs. predicted (Hybrid ARIMA-WBF
Model) values for South Korea COVID-19 data; (b) Real-time forecasts
(10 days) of the number of cases for South
Korea}\label{fit_forecast_skorea}
\end{figure}

\begin{figure}[H]
\centering
\includegraphics[width=0.45\textwidth]{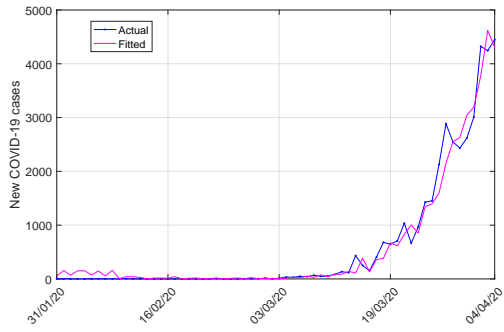}(a)
\includegraphics[width=0.45\textwidth]{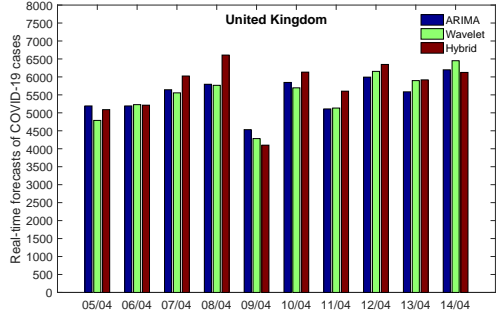}(b)
\caption{Figures of (a) Actual Vs. predicted (Hybrid ARIMA-WBF
Model) values for the UK COVID-19 data; (b) Real-time forecasts (10
days) of the number of cases for the UK}\label{fit_forecast_uk}
\end{figure}

It is interesting to see that the error series (residuals) generated by ARIMA are oscillating and nonstationary for all the datasets. These seasonal oscillations can be captured through the wavelet transform, which can decompose a time series into a linear combination of different frequencies. These residual series as in Figure \ref{residuals}) satisfy the admissibility condition (zero mean) that forces wavelet functions to wiggle (oscillate between positive and negative), a typical property of wavelets. Thus, we remodel the residuals obtained using the ARIMA model with that of the WBF model.  The value of Wavelet levels is obtained by using the formula, as mentioned in Algorithm \ref{algo1}. WBF model was implemented using `\textit{WaveletArima}' \citep{paul2017package} package in R software with `periodic' boundary and all the other parameters were kept as default. As the WBF model is fitted on the residual time series, predictions are generated for the next ten time steps (5 April 2020 to 14 April 2020). Further, both the ARIMA forecasts and WBF residual forecasts are added together to get the final out-of-sample forecasts for the next ten days (5 April 2020 to 14 April 2020). The hybrid model fittings (training data) for five countries, namely Canada, France, India, South Korea and the UK are displayed in Figures \ref{fit_forecast_canada}(a), \ref{fit_forecast_france}(a), \ref{fit_forecast_india}(a), \ref{fit_forecast_skorea}(a) and \ref{fit_forecast_uk}(a) respectively. The real-time (short-term) forecasts using ARIMA, WBF, and hybrid ARIMA-WBF model for Canada, France, India, South Korea, and the UK are displayed in Figures \ref{fit_forecast_canada}(b),
\ref{fit_forecast_france}(b), \ref{fit_forecast_india}(b),
\ref{fit_forecast_skorea}(b) and \ref{fit_forecast_uk}(b)
respectively.

The predicted values for the training COVID-19 cases datasets of the proposed hybrid model for five countries are further used for model adequacy checking and based on actual and predicted test outputs, we computed RMSE and MAE for all the datasets and reported them in Table \ref{table_metrics_rmse}. The performances of the proposed hybrid ARIMA-WBF model are superior as compared to the individual models for Canada, France, and the UK, whereas, for India and South Korea, our results are competitive with ARIMA. It is often true that no model can be universally employed in all circumstances, and this is in relevance with ``no free lunch theorem" \citep{wolpert1997no}. Even if in a very few cases hybrid ARIMA-WBF model gave lower information criteria values (in terms of RMSE and MAE for training data), we still can opt for the hybrid model given the asymmetric risks involved as we believe that it is better to take decisions based on a hybrid model rather than depending on a single one at least for this pandemic. We produced ten-days-ahead point forecasts based on all the three model models discussed in this chapter and reported then in Figures \ref{fit_forecast_canada}-\ref{fit_forecast_uk}. Our model can easily be updated on a daily or periodic basis once the actual values are received for the country-wise COVID-19 cases. 

\begin{table}[H]
\centering \small\caption{RMSE and MAE values for different
forecasting models on five time series (training data only) data
sets}\mbox{
   \begin{tabular}{|c|c|c|c|c|}
        \hline
        \multirow{2}{*}{Model}       & Performance  & ARIMA    & WBF    & Hybrid     \\
                                     & Metrics      &          &        & ARIMA-WBF Model \\ \hline
        \multirow{2}{*}{Canada}      & RMSE         & 150.05   & 202.64 & 149.60  \\
                                     & MAE          & 41.68    & 89.21  & 40.05   \\ \hline
        \multirow{2}{*}{France}      & RMSE         & 710.46   & 740.06 & 631.91  \\
                                     & MAE          & 358.87   & 441.97 & 306.78  \\ \hline
        \multirow{2}{*}{India}       & RMSE         & 50.83    & 68.38  & 55.25   \\
                                     & MAE          & 16.07    & 31.78  & 24.00  \\ \hline
        \multirow{2}{*}{South Korea} & RMSE         & 81.81    & 82.78  & 90.29   \\
                                     & MAE          & 44.71    & 47.81  & 54.06   \\ \hline
       \multirow{2}{*}{UK}           & RMSE         & 209.36   & 405.87 & 180.66  \\
                                     & MAE          & 104.28   & 248.83 & 100.68  \\ \hline
    \end{tabular}}
    \label{table_metrics_rmse}\label{table2}
\end{table}

\begin{remark}
Please note that this is not an ex-post analysis, but a real, live forecasting exercise. Thus, these real-time short-term forecasts based on the proposed hybrid ARIMA-WBF model for Canada, France, India, South Korea, and the UK will be helpful for government officials and policymakers to allocate adequate health care resources for the coming days.
\end{remark}

\section{Risk Assessment of COVID-19 cases}\label{section_3}

At the outset of the COVID-19 outbreak, data on country-wise case fatality rates due to COVID-19 were obtained for 50 affected countries. The case fatality rate can be crudely defined as the number of deaths in persons who tested positive for divided by the number of COVID-19 cases. In this section, we are going to find out a list of essential causal variables that have strong influences on the CFR. The datasets and codes of this section are made publicly available at \url{https://github.com/indrajitg-r/COVID} for the reproducibility of this work. 

\subsection{Data}

In the face of rapidly changing data for COVID-19, we calculated the case fatality ratio estimates for 50 countries from the day of starting the outbreak to 4 April 2020 from the following website\footnote{\url{https://www.worldometers.info/coronavirus/}}. A lot of preliminary analysis is done to determine a set of possible variables, some of which are expected to be critical causal variables for risk assessments of COVID-19 in these affected countries. Previous studies \citep{nishiura2020serial, russell2020estimating, jung2020real, boldog2020risk} have suggested that the total number of cases, age distributions, and shutdown period have high impacts on the CFR values for some of the countries. Along with these three variables, we also considered seven more demographic structures and disease characteristics for these countries as input variables that are likely to have a potential impact on the CFR estimates. Therefore, the CFR modeling dataset consists of 50 observations having ten possible causal variables and one numerical output variable (viz. CFR), as reported in Table \ref{table3}.

\begin{table}[H]
\centering \small \caption{Descriptive statistics of possible causal variables and the response variable of CFR dataset for 50 countries.}
\makebox[\textwidth]{%
    \begin{tabular}{|c|c|c|c|c|c|c|}
        \hline
        {Input and Output variables}      & Notation & Variable Type      & Mean   & Variance  & Min. Value & Max. Value  \\ \hline
        {Total cases (in thousands)}      & x.x1     & Numerical          & 20.89  & 2187.92   & 0.25      & 277.96       \\
        {population (in millions)}        & x.x2     & Numerical          & 110.62 & 73658.12  & 0.03      & 1402.01      \\
        {population density per km$^2$}   & x.x3     & Numerical          & 139.78 & 20371.56  & 3.00      & 568          \\
        {\% people $>$ 65 years age}      & x.x4     & Numerical          & 13.58  & 38.59     & 3.20      & 27           \\
        {lockdown days count}             & x.x5     & Numerical          & 20.20  & 95.96     & 0         & 73           \\
        {time period (in days)}           & x.x6     & Numerical          & 48.72  & 309.23    & 25        & 84           \\
        {doctors per 1000 people}         & x.x7     & Numerical          & 2.71   & 1.98      & 0.20      & 6.36         \\
        {Hospital beds per 1000 persons}  & x.x8     & Numerical          & 3.92   & 8.24      & 0.10      & 13.70        \\
        {Income standards}                & x.x9     & Categorical        & -      & -         & 0         & 1            \\
        {Climate zones}                   & x.x10    & Categorical        & -      & -         & -1        & 1            \\
        {CFR (response variable)}         & Y        & Numerical          & 0.041  & 0.001     & 0.005     & 0.127
        \\ \hline
    \end{tabular}\label{table3}
    }
\end{table}

The possible causal variables considered in this study are the followings: the total number of COVID-19 cases (in thousands) in the country till 4 April, 2020, population density per km$^2$ for the country, total population (in millions) of the country (approx.), percentage of people in the age group of greater than 65 years, lockdown days count (from the starting day of lockdown till April 4, 2020), time-period (in days) of COVID-19 cases for the country (starting date to April 4, 2020), doctors per 1000 people in the country, hospital beds per 1000 people in the country, income standard (e.g., high or lower) of the country and climate zones (e.g., tropical, subtropical or moderate) of the country. The dataset contains a total of 8 numerical input variables and two categorical input variables. 

\subsection{Method: Regression Tree}

For the risk assessment with the CFR dataset for 50 countries, we apply the regression tree (RT) \citep{breiman1984classification} that has built-in feature selection mechanism, easy interpretability, and provides better visualization. Rt, as a widely used simple machine learning algorithm, can model arbitrary decision boundaries. The methodology outlined in \citep{breiman1984classification} can be summarized into three stages. The first stage involves growing the tree using a recursive partitioning technique to select essential variables from a set of possible causal variables and split points using a splitting criterion. The standard splitting criteria for RT is the mean squared error (MSE). After a large tree is identified, the second stage of RT methodology uses a pruning procedure that gives a nested subset of trees starting from the largest tree grown and continuing the process until only one node of the tree remains. The cross-validation technique is popularly used to provide estimates of future prediction errors for each subtree. The last stage of the RT methodology selects the optimal tree that corresponds to a tree yielding the lowest cross-validated or testing set error rate. To avoid instability of trees in this stage, trees with smaller sizes, but comparable in terms of accuracy, are chosen as an alternative. This process can be tuned to obtain trees of varying sizes and complexity. A measure of variable importance can be achieved by observing the drop in the error rate when another variable is used instead of the primary split. In general, the more frequent a variable appears as a primary split, the higher the importance score assigned. A detailed description of the tree building process is available at \citep{james2013introduction}.

\subsection{Results}

\begin{figure}[H]
\centering
\includegraphics[width=\textwidth]{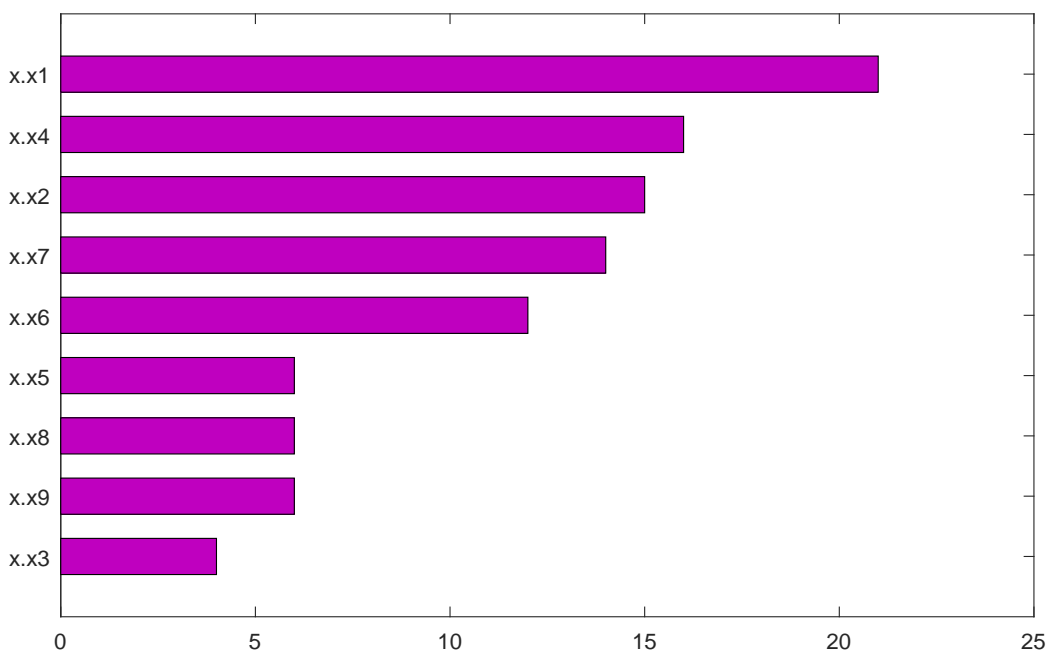}
\caption{Variable Importance Percentages affecting the CFR based on
a complexity parameter in RT} \label{imp}
\end{figure}

\begin{figure}[H]
\centering
\includegraphics[width=\textwidth]{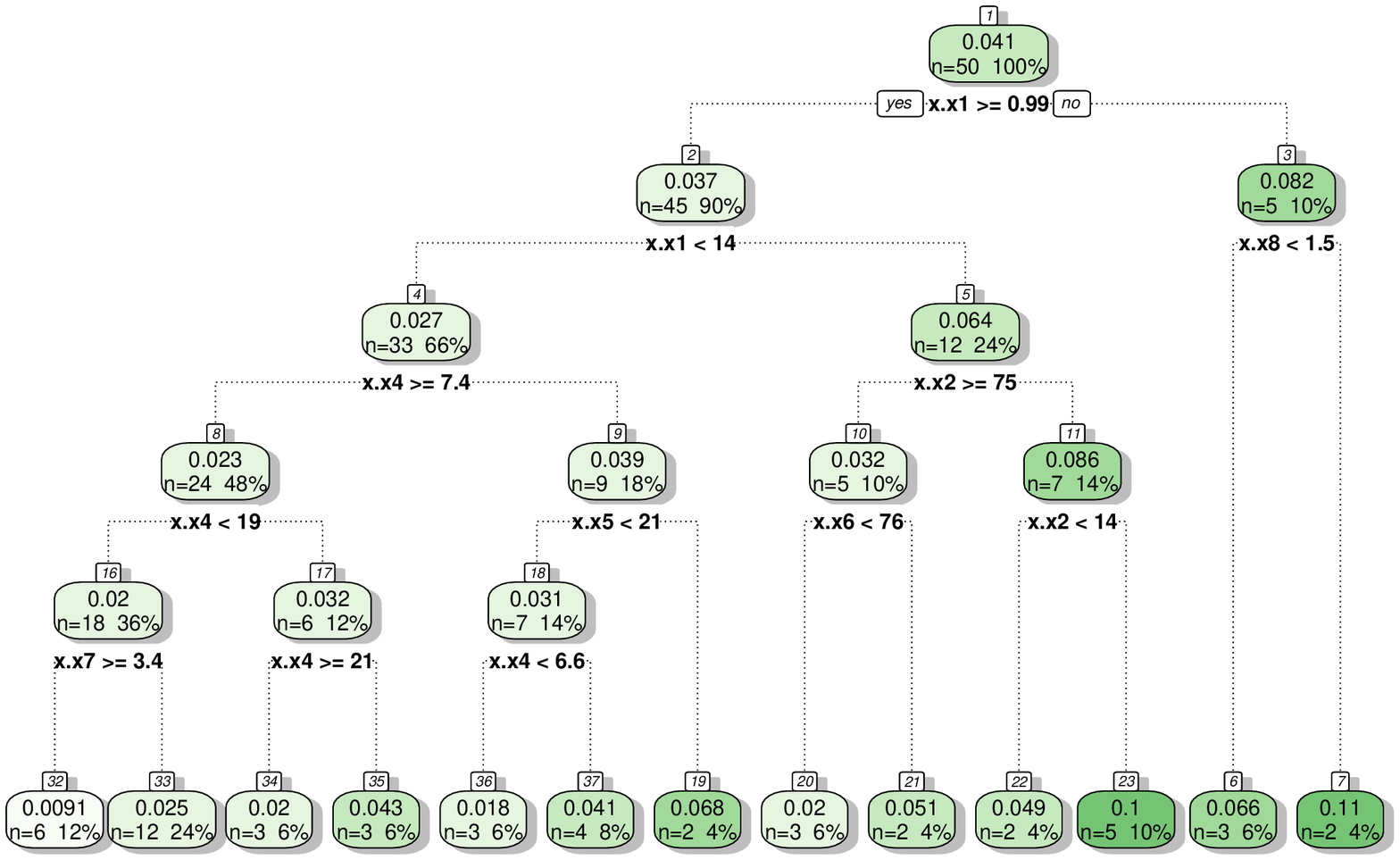}
\caption{Optimal tree representing the relationships between the causal variables and CFR} \label{tree}
\end{figure}

The rationale behind the choice of RT as a potential model to find the important casual variables out of 10 input variables for the CFR estimates is the simplicity, easy interpretability, and high accuracy of the RT algorithm.  We apply an optimal RT model to the dataset consisting of 50 different country samples and try to find out potential casual variables from the set of available variables that are related to the case-fatality rates. RT is implemented using `\textit{rpart}' \citep{therneau2015package} package in R with ``minsplit" equals to 10\% of the data as a control parameter. We have used RMSE, co-efficient of multiple determination $(R^{2})$, and adjusted $R^{2}$ ($Adj R^{2}$) to evaluate the predictive performance of the tree model used in this study \citep{james2013introduction}. An optimal regression tree is built with 7 variables with `minsplit' = 5 with equal costs for each variable. The estimates of the performance metrics for the fitted tree are as follows:
RMSE = 0.013, $R^{2} = 0.896$, and $Adj R^{2}= 0.769$. A variable importance list from the RT is given in Figure \ref{imp} and the fitted tree is provided in Figure \ref{tree}. 

From the variable importance plot based on the complexity parameter of the RT model (also see Figure \ref{imp}), seven causal variables are obtained out of 10 potential input variables having higher importance. These seven causal variables that significantly affect the CFR for 50 most affected countries are the followings: total number of COVID-19 cases in the country (in thousands), percentage of people in the age group of greater than 65 years, total population (in millions) of the country, doctors per 1000 people in the country, lockdown period (in days) for the country, time-period (in days) of COVID-19 cases for the country, and hospital beds per 1000 people in the country. Our results are consistent with previous results obtained by \citep{russell2020estimating, jung2020real, boldog2020risk}, where the authors suggested that the total number of cases, age distributions, and shutdown period have high impacts on the CFR estimates. But interestingly, we obtained four more essential causal variables that will provide some new insights into the study of risk assessments for COVID-19 affected countries. Out of these 7 numerical input variables, there are four control variables (number of cases, people of age group $>$ 65 years, lockdown period, and hospital beds per 1000 people) present that can be managed to fight against this deadly disease. Once these variables are taken care of, the respective country may reduce their case fatality rate at a significant rate. 

Figure \ref{tree} shows the relationship between the important causal variables and CFR. In Figure \ref{tree}, the tree starts with the total number of COVID-19 cases as the most crucial causal variable in the parent node. In each box, the top most numerical values suggest the average CFR estimates based on the tree. One of the key findings of the tree is the following rule: When the number of cases of a country is greater than 14,000 having a population between 14 to 75 million are having second highest case fatality rate, viz., 10\%. Similarly, one can see all the rules generated by RT to get additional information about the relationships between control parameters and the response CFR variable. 

\section{Limitations of our findings}\label{section_4}

We made some simplifying assumptions to carry out the analysis of
COVID-19 datasets. The assumptions are listed as follows: (a) the
virus mutation rate are comparable for different countries; (b) the
recovered persons will achieve permanent immunity against COVID-19; (c) we ignore the effect of climate change (also spatial data structures) during the short-term predictions. Along in this line, we presented two different
approaches to deal with two inter-connected problems on COVID-19. In the first problem of short-term predictions for COVID-19 outbreak in five countries, we proposed a hybrid methodology combining ARIMA and WBF models. In the second problem of risk assessment, we found some important factors affecting case fatality rates of COVID for 50 highly affected nations. However, there may exist a few more controllable factor(s), and some disease-based characteristics that can also have an impact on the value of CFR for different countries, can be regarded as future scope of the study.

\section{Discussions}\label{section_5}

The COVID-19 outbreaks globally present a significant challenge for modelers, as there are limited data available on the early growth
trajectory, and epidemiological characteristics of the novel
coronavirus have not been fully elucidated. In this study, we
considered two alarmingly important problems relevant to ongoing
COVID-19 pandemic. The first problem deals with the real-time
forecasts of the daily COVID-19 cases in five different countries.
We proposed a hybrid ARIMA-WBF model that can explain the nonlinear and nonstationary behavior present in the univariate time series datasets of COVID-19 cases. Ten days-ahead forecasts are provided for Canada, France, India, South Korea, and the UK. The proposed model can be used as an early warning system to fight against the COVID-19 pandemic. Below we present a list of suggestions based on the results of the real-time forecasts.
\begin{enumerate}
\item Since we presented a real-time forecast system unlike an ex-post analysis, thus one can regularly update the actual confirmed cases and update the predictions, just like it happens in weather forecasting. 
\item The forecasts mostly show oscillating behavior for the next 10 days and reflect the impact of the broad spectrum of social distancing measures implemented by the governments, which likely helped stabilize the epidemic. 
\item The short-term forecasts don't necessarily show any stiff decay sooner; also, these five countries are not going to face any unlike uplifts in the number of cases too.
\item Guided by the short-term forecasts reported in this paper, the lockdown period can be adjusted accordingly. 
\end{enumerate}

Secondly, we assessed the risk of COVID-19 by finding seven key
parameters that are expected to have powerful associations with
that of case fatality rates. This is done by designing an optimal
regression tree model, a simplified machine learning approach. The
model is very flexible, easily interpretable, and the more data will
come, one can just incorporate the new data sets and rebuild the
trees to get the updated estimates. RT provides a better visual
representation and is easily interpretable to be understood by a
broader audience. Quantification of the outbreak risks and their dependencies on the key parameters will support the governments and policymakers for the planning of health care systems in different countries that faced this epidemic. Experimental results suggest four control variables out of seven highly influential variables that will have a significant impact on controlling CFR. Below we present a point by point discussion of the control variables affecting CFR and preventive actions to be taken by the governments.
\begin{enumerate}
\item The number of covid cases of the country can be reduced by enforcing social distancing strategies.
\item Number of people of age group $>$ 65 years should be specially taken care of and isolated.
\item Lockdown time period can be extended if the country faces a sharp increase in the number of cases and or deaths.
\item The number of hospital beds should be increased by making special health care arrangements in other places to deal with this emergency due to COVID-19.
\end{enumerate} 

%\section*{Acknowledgements}
%The authors are grateful to the editors and anonymous referees for
%careful reading, constructive comments and insightful suggestions,
%which have greatly improved the quality of the paper.

\bibliographystyle{plain}
\biboptions{square}
\bibliography{bibliography}
\end{document}